\documentclass{elsart}
\usepackage[english]{babel}
\usepackage[OT1,T1]{fontenc}
\usepackage[latin1]{inputenc}
\usepackage{amsmath}
\usepackage{natbib}
\usepackage[colorlinks]{hyperref}
\usepackage{hypernat}

\usepackage{graphicx}
\usepackage{mathrsfs}
\usepackage{amssymb}
\usepackage{amsfonts}
\usepackage{dsfont}
\usepackage{latexsym}
\newtheorem{theorem}{Theorem}[section]
\newtheorem{lemma}[theorem]{Lemma}
\newtheorem{proposition}[theorem]{Proposition}
\newtheorem{corollary}[theorem]{Corollary}

\newtheorem{remark}{Remark}

\newcommand{\I}{\mathds{1}}

\begin{document}

\begin{frontmatter}
\title{A quantile-copula approach to conditional density estimation.}
%\protect\thanksref{T1}}
%\runtitle{Conditional density estimation}
%\thankstext{T1}{Footnote to the title with the `thankstext' command.}

%%\begin{aug}
\author{Olivier P. Faugeras }
\ead{olivier.faugeras@gmail.com}
%\author{\fnms{Second} \snm{Author}\thanksref{t3}\ead[label=e2]{second@somewhere.com}}
%\and
%\author{\fnms{Third} \snm{Author}\thanksref{t1}
%\ead[label=e3]{third@somewhere.com}
%\ead[label=u1,url]{http://www.foo.com}}

%\thankstext{t1}{Some comment}
%\thankstext{t2}{First supporter of the project}
%\thankstext{t3}{Second supporter of the project}
\runauthor{Olivier Faugeras}
%\journal{Journal of Multivariate Analysis}
%\affiliation{L.S.T.A, Université Paris 6}

\address{ L.S.T.A, Université Paris 6\\175, rue du Chevaleret, 75013 Paris, France \\
Tel:+(33) 1 44 27 85 62\\ Fax:+(33) 1 44 27 33 42}
%\printead{e1}\\
%\phantom{E-mail:\ }}

%\end{aug}

\begin{abstract}
We present a new non-parametric estimator of the conditional density of the kernel type. It is based on an efficient transformation of the data by quantile transform. By use of the copula representation, it turns out to have a remarkable product form.  We study its asymptotic properties and compare its bias and variance to competitors based on nonparametric regression. A comparative numerical simulation is provided.
\end{abstract}


\end{frontmatter}

%--------------------------------------------------

\section{Introduction}

\subsection{Motivation}
Let $( (X_i,Y_i);i=1,\ldots,n )$ be an independent identically distributed sample from real-valued random variables $(X,Y)$ sitting on a given probability space. For predicting the response $Y$ of the input variable $X$ at a given location $x$, it is of great interest of estimating not only the conditional mean or \emph{regression function} $E(Y|X=x)$, but the full \emph{conditional density} $f(y|x)$. Indeed, estimating the conditional density is much more informative, since it allows not only to recalculate the conditional expected value $E(Y|X)$ and conditional variance from the density, but also to provide the general shape of the conditional density.
 This is especially important for multi-modal or skewed densities, which often arise from nonlinear or non-Gaussian phenomenas, where
the expected value might be nowhere near a mode, i.e. the most likely value to appear.
% Therefore, considering the expected value as the best predictor, (which is the case from a mathematical standpoint for a decision based on the choice, yet arbitrary, of the $\mathbb{L}_2$ norm) is questionable. 
Moreover, for situations in which
confidence intervals are preferred to point estimates, the estimated conditional density is an object of obvious interest.

\subsection{Estimation by kernel smoothing}
A natural approach to estimate the conditional density $f(y|x)$ of $Y$ given $X=x$ would be to exploit the identity 
\begin{align}
f(y|x) = \frac{{f_{XY} (x,y)}}{{f_X (x)}} \label{ratio}
\end{align}
where $f_{XY}$ and $f_X$ denote the joint density of $(X,Y)$ and $X$, respectively. By introducing Parzen-Rosenblatt kernel estimators of these densities, namely 
\begin{align*}
 \hat f_{n,XY} (x,y): &= \frac{1}{n}\sum\limits_{i = 1}^n {K'_{h'} (X_i  - x)} K_h (Y_i  - y) \\ 
 \hat f_{n,X} (x): &= \frac{1}{n}\sum\limits_{i = 1}^n {K'_{h'} (X_i  - x)}  
 \end{align*}
where $K_h(.)=1/hK(./h)$ and $K'_{h'}(.)=1/h'K'(./h')$ are (rescaled) kernels with their associated sequence of bandwidth $h=h_n$ and $h'=h'_n$ going to zero as $n\rightarrow \infty$, one can construct the quotient \[
\hat f_n^R(y|x): = \frac{{\hat f_{n,XY} (x,y)}}{{\hat f_{n,X} (x)}}
\]
and obtain an estimator of the conditional density. Such an estimator was first studied by Rosenblatt \cite{R1969}, and more recently by Hyndman et al. \cite{HBG1996}, who slightly improved on Rosenblatt's kernel based estimator. 
%See also the book of Ferraty and Vieu \cite{FV2006} for an extension to functional data. 

\subsection{Estimation by regression techniques}
As pointed out by numerous authors, see e.g. Fan and Yao \cite{FY2005} chapter 6, this approach is equivalent to the one arising from considering this conditional density estimation problem in a regression framework. Indeed, let $F(y|x)$ be the cumulative conditional distribution function of $Y$ given $X=x$. It stems  from the fact that \[
E\left( {\I_{\left| {Y - y} \right| \le h} |X = x} \right) = F(y + h|x) - F(y - h|x) \approx 2h.f(y|x)
\]
as $h\rightarrow 0$, that, if one replace the expectation in the above expression by its empirical counterpart, one can apply the usual local averaging methods and perform a regression estimation on the synthetic data $((1/2h)\I_{\left| {Y_i - y} \right| \le h}$; $i=1,\ldots,n)$. By a Bochner type theorem, one can even replace the transformed data by its smoothed version  $$Y_i':=
K_h (Y_i - y): = \frac{1}{h}K\left( {\frac{{Y_i - y}}{h}} \right).
$$ 
In particular,  the popular Nadaraya-Watson regression estimator \[
\hat f_n^{NW} (y|x) := \frac{{\sum\nolimits_{i = 1}^n {Y'_i K'_{h'} (X_i  - x)} }}{{\sum\nolimits_{i = 1}^n {K'_{h'} (X_i  - x)} }}
\]
 reduces itself to the same estimator of the conditional density of the double kernel type as before  $$\hat f_n^{NW}(y|x):= \frac{{\sum\nolimits_{i = 1}^n {K_h (Y_i - y) K'_{h'} (X_i  - x)} }}{{\sum\nolimits_{i = 1}^n {K'_{h'} (X_i  - x)} }}=\hat f_n^R(y|x).$$
 
 Taking advantage of this regression formulation, Fan, Yao and Tong \cite{FYT1996} proposed a conditional density estimator which generalizes the kernel one by use of the local polynomial techniques. In particular, it allows to tackle with the bias issues of the kernel smoothing. However, and unlike the former, it is no longer guaranteed to have positive value nor to integrate to 1 with respect to $y$. With these issues in mind, Hyndman and Yao \cite{HY2002} built on local polynomial techniques and suggested two improved methods, the first one based on locally fitting a log-linear model and the second one on constrained local polynomial modeling.  An overview can be found in Fan and Yao \cite{FY2005} (chapter 6 and 10). Very recently, Györfi and Kohler \cite{GK2007} studied a partitioning type estimate and studied its properties in total variation norm and Lacour \cite{L2007} a projection-type estimate for Markov chains. 
 
\subsection{A product shaped estimator}
However, these two equivalent approaches suffer from several drawbacks: first, by its form as a quotient of two estimators, the probabilistic behavior of the Nadaraya-Watson estimator (or its local polynomial counterpart) is tricky to study. It is usually dealt with by a centering at expectation for both numerator and denominator and a linearizing of the inverse, see e.g. \cite{FY2005}, or \cite{B1998} for details. Second, at a conceptual level, one could argue that implementing regression estimation techniques in this setting is, in a sense, unnatural: estimating a  density, even if it is a conditional one, should resort to density estimation techniques only.  Finally, practical implementations of these estimators can lead to numerical instability when the denominator is close to zero.

To remedy these problems, we propose an estimator which builds on the idea of using synthetic data, i.e. a representation of the data more adapted to the problem than the original one. By transforming the data by quantile transforms and making use of the copula function, the estimator turns out to have a remarkable \emph{product} form
$$\hat f_n(y|x)=\hat f_Y(y) \hat c_n(F_n(x),G_n(y))$$ 
where $\hat f_Y$,  $\hat c_n$, $F_n(x)$, $G_n(y)$ are estimators of the density $f_Y$ of $Y$, the copula density $c$, the c.d.f. $F$ of $X$ and $G$ of $Y$ respectively  (see next section below for definitions).
Its study then reveals to be particularly simple: it reduces to the ones already done on nonparametric density estimation.  
  
  The rest of the paper is organized as follows: %in the remaining of this section, a brief overview of the literature is sketched. 
 in section 2, we introduce the quantile transform and the copula representation which leads to the definition of our estimator. In section 3, the main asymptotic results are established and compared in section 4 to those of  other competitors. Proofs are mainly based on a series of auxiliary lemmas which are given in section 5.

%---------------------------------------------------------------------------------------------------------
%---------------------------------------------------------------------------------------------------------
 
\section{Presentation of the estimator}
For sake of simplicity and clarity of exposition, we limit ourselves to unidimensional real valued input variables $X$. However, all the results of this article can be easily extended to the multivariate case. 
\subsection{The quantile transform}
The idea of transforming the data is not new. It has been used to improve the range of applicability and performance of classical estimation techniques, e.g. to deal with skewed data, heavy tails, or restrictions on the support (see e.g. Devroye and Lugosi \cite{DL2001} chapter 14 and the references therein, and also Van der Vaart \cite{VDV1998} chapter 3.2 for the related topic of variance stabilizing transformations in a parametric context). In order to make inference on $Y$ from $X$, a natural question which then arises is, what is the ``best''  transformation, if this question has a sense. As one can note from the above references, the ``best'' transformation is very linked to the distribution of the underlying data.  We will see below that, for our problem, the natural candidate is  the quantile transform.

The quantile transform is a well-known probabilistic trick which is used  to reduce proofs, e.g. in empirical process theory, for arbitrary real valued random variables $X$ to ones for random variables $U$ uniformly distributed on the interval $[0,1]$. It is based on the following well-known fact that whenever $F$ is continuous, the random variable $U=F(X)$ is uniformly distributed on $(0,1)$ and that conversely, when $F$ is arbitrary, if $U$ is a uniformly distributed random variable on $(0,1)$, $X$ is equal in law to $F^{-1}(U)$, where $F^{-1}=Q$ is the generalized inverse or quantile function of $X$. (See e.g. \cite{SW1986}, chapter 1).

As a consequence, given a sample $(X_1, \ldots , X_n)$ of random variables with common continuous c.d.f. $F$ sitting on a probability space $(\Omega,\mathcal{A},\mathbb{P})$, one can always enlarge this probability space to carry a sequence $(U_1, \ldots ,U_n)$ of uniform $(0,1)$ random variables such that $U_i=F(X_i)$, that is to say to construct a pseudo-sample with a \emph{prescribed uniform} marginal distribution. 

%----------------------------------------------------
\subsection{The copula representation}
Formally, a copula is a bi-(or multi)variate distribution function whose margi\-nal distribution functions are uniform on the interval $[0,1]$. Indeed, %solving a problem formulated by Fréchet \cite{F1951},
 Sklar \cite{S1959} proved the following fundamental result:
\begin{theorem} \label {sklar}
For any bivariate cumulative distribution function $F_{X,Y}$ on $\mathds{R}^2$, with marginal cumulative distribution functions $F$ of $X$ and $G$ of $Y$, there exists some function $C:[0,1]^2\rightarrow [0,1]$, called the dependence or copula function, such as 
\begin{align}
F_{X,Y}(x,y)=C(F(x),G(y)) \mbox{ , } -\infty \le x,y\le +\infty .\label{skl}
\end{align}
If $F$ and $G$ are continuous, this representation is unique with respect to $(F,G)$. The copula function $C$ is itself a cumulative distribution function on $[0,1]^2$ with uniform marginals.
\end{theorem}
 
This theorem gives a representation of the bivariate c.d.f. as a function of each univariate c.d.f. In other words, the copula function captures the dependence structure among the components $X$ and $Y$ of the vector $(X,Y)$, irrespectively of the marginal distribution $F$ and $G$. Simply put, it allows to deal with the randomness of the dependence structure and the randomness of the marginals \emph{separately}.

 Copulas appears to be naturally linked with the quantile transform as formula \ref{skl} entails that $C(u,v)=F_{X,Y}(F^{-1}(u),G^{-1}(v))$.
For more details regarding copulas and their properties, one can consult for example the book of Joe \cite{J1997}. Copulas have witnessed a renewed interest in statistics, especially in finance, since the pioneering work of Deheuvels \cite{D1979}, who introduced the empirical copula process. Weak convergence of the empirical copula process was investigated by Deheuvels \cite{D1981}, Van der Vaart and Wellner \cite{VDVW1996}, Fermanian, Radulovic and Wegkamp \cite{FRW2004}. For the estimation of the copula density, refer to Gijbels and Mielniczuk \cite{GM1990},   Fermanian \cite{F2005} and Fermanian and Scaillet \cite{FS2003}. 

From now on, we assume that the copula function $C(u,v)$ has a density $c(u,v)$ with respect to the Lebesgue measure on $[0,1]^2$ and that $F$ and $G$ are strictly increasing and differentiable with densities $f$ and $g$. $C(u,v)$ and $c(u,v)$ are then the cumulative distribution function (c.d.f.) and density respectively of the transformed variables $(U,V)=(F(X),G(Y))$. By differentiating formula \eqref{skl}, we get for the joint density,
   $$f_{XY} (x,y) = \frac{{\partial ^2 F_{XY} (x,y)}}{{\partial x\partial y}} = f (x)g (y)c(F (x),G (y))$$
   where $c(u,v) := \frac{{\partial ^2 C(u,v)}}{{\partial u\partial v}}$ is the above mentioned copula density. Eventually, we can obtain the following  explicit formula of the conditional density
\begin{align}\label{densitecondi}
	f_{Y|X} (x,y) = \frac{{f_{XY} (x,y)}}{{f (x)}} = g (y)c(F (x),G (y)) .
\end{align}

%-------------------------------------------------------------------------------
%----------------------------------------------------------------------------
\subsection{Construction of the estimator}
%From now on, we simplify notations and note $f$ and $F$ the density and c.d.f. of $X$, and $g$ and $G$ those of $Y$.
 Starting from the previously stated product type formula \eqref{densitecondi}, 
a natural plug-in approach to build an estimator of the conditional density is to use 
\begin{itemize}
	\item a Parzen-Rosenblatt kernel type non parametric estimator of the marginal density $g$ of $Y$, 
\[
\hat g_n (y) := \frac{1}{nh_n}\sum\limits_{i = 1}^n {K_0\left( {\frac{{y - Y_i }}{{h_n }}} \right)} 
\]
  \item the empirical distribution functions $F_n (x)$ and $G_n (y)$ for $F(x)$ and $G(y)$ respectively,
    $$F_n (x) = \frac{1}{n}\sum\limits_{j = 1}^n \I_{X_j  \le x} \mbox{ and }
 G_n (y) := \frac{1}{n}\sum\limits_{j = 1}^n {\I_{Y_j  \le y} } .$$    
\end {itemize}   

Concerning the copula density $c(u,v)$, we noted that $c(u,v)$ is the joint density of the transformed variables $(U,V)=(F(X),G(Y))$. Therefore, $c(u,v)$ can be estimated by the  bivariate Parzen-Rosenblatt kernel type non parametric density (pseudo) estimator,
\begin{align}\label{cn}
c_n (u,v) := \frac{1}{na_nb_n}\sum\limits_{i = 1}^n {K\left( {\frac{{u - U_i }}{{a_n }},\frac{v-V_i}{b_n}} \right)} 
\end{align}
where $K$ is a bivariate kernel and $a_n$, $b_n$ its associated bandwidth. For simplicity, we restrict ourselves to product kernels, i.e. $K(u,v)=K_1(u)K_2(v)$ with the same bandwidths $a_n=b_n$.

Nonetheless, since $F$ and $G$ are unknown, the random variables $(U_i,V_i)$ are not observable, i.e. $c_n$ is not a true statistic. Therefore, we approximate the pseudo-sample $(U_i,V_i), i=1, \ldots, n$ by its empirical counterpart $(F_n(X_i),G_n(Y_i)), i=1, \ldots, n$. 
We therefore obtain a genuine estimator of $c(u,v)$
\begin{align} \label{hatcn}
\hat c_n (u,v) := \frac{1}{na_n^2}\sum\limits_{i = 1}^n {K_1 \left( {\frac{{u - F_n (X_i )}}{{a_n }}} \right)K_2 \left( {\frac{{v - G_n (Y_i )}}{{a_n }}} \right)} .
\end{align}

Eventually, the conditional density estimator is written as
\begin{align*}
   \hat f_n (y|x) :=& \left[ {\frac{1}{nh_n}\sum\limits_{i = 1}^n {K_0\left( {\frac{{y - Y_i }}{{h_n }}} \right)} } \right].\left[ {\frac{1}{na_n^2}\sum\limits_{i = 1}^n {K_1 \left( {\frac{{F_n (x) - F_n (X_i )}}{{a_n }}} \right)} } \right.\\
   &{ } \left. {K_2 \left( {\frac{{F_n (y) - G_n (Y_i )}}{{a_n }}} \right)} \right]
\end{align*}
or, under a more compact form, 
\begin{align}
	\hat f_n (y|x) := \hat g_n (y)\hat c_n (F_n (x),G_n (y)) \label{estimateur}.
\end{align}

\begin{remark}
To our knowledge, the estimator studied in this paper has never been proposed in the literature. However, some connections can be made with the nearest neighbor one proposed by  Stute \cite{S1984}, \cite{S1986a} and \cite{S1986b} for conditional cumulative distribution function and the Gasser and Müller \cite{GM1979} and Priestley and Chao \cite{PC1972} one in the context of regression estimation. Indeed, these estimators  tackle the issue of having a random denominator  by first transforming the design $X_1,\ldots,X_n$ to a uniform (random) one. This result in assigning the surfaces under the kernel function instead of its heights as weights. Contrary to our estimator, they do not make transformations of the data in both directions $X$ and $Y$. 
% The Gasser and Müller estimator, which, , has a convolution shape	as shown below:
%\begin{align*}
%m_n^{GM(1)}  &:= \frac{1}
%{{h_n }}\sum\limits_{i = 1}^{n - 1} {\left\{ {\int_{X_{i,n} }^{X_{i + 1,n} } {K\left( {\frac{{x - u}}
%{{h_n }}} \right)du} } \right\}} Y_{\left[ i \right]} \\
%m_n^{GM(2)}  &:= \frac{1}
%{{h_n }}\sum\limits_{i = 1}^n {(X_{i + 1,n}  - X_{i,n} )K\left( {\frac{{x - X_{i,n} }}
%{{h_n }}} \right)Y_{[i]} } 
%\end{align*}
%where $X_{i,n}$ denotes the ith order statistic of the sample $(X_1,\ldots,X_n)$ and $Y_{[i]}$ its corresponding $Y$ value.
\end{remark}
%
%\begin{remark}
% Instead of a product kernel, we could have also used a semi norm-based kernel, as advocated, e. g. by \cite{FV2006}. Interested readers should report to the above cited reference. 
%\end{remark}

%----------------------------------------------------------------------------
%---------------------------------------------------------------------------

\section{Asymptotic results}

\subsection{Notations and assumptions}
We note the ith moment of a generic kernel (possibly multivariate)  $K$ as $m_i (K) := \int {u^i K(u)du} $, and the $\mathbb{L}_p$ norm of a function $h$ by $||h||_p:=\int {h^p}$. We use the sign $\simeq$ to denote the order of the bandwidths, i.e. $h_n \simeq u_n$ means that $h_n=c_nu_n$ with $c_n \to c>0$. The support of the densities function $f$ and $c$ are noted as
$\mbox{supp}(f)=\{x\in \mathds{R};f(x)>0\}$ and 
$\mbox{supp}(c)=\{(u,v)\in \mathds{R}^2;c(u,v)>0\}$, respectively.

For stating our results, we will have to make some regularity assumptions on the kernels and the densities which, although far from being minimal, are somehow customary in kernel density estimation (see subsection \ref{convdens1} for discussions and details). Set $x$ and $y$ two fixed points in the interior of $\mbox{supp}(f)$ and $\mbox{supp}(g)$ respectively. In the remainder of this paper, we will always suppose that
\begin{enumerate}
	\item [i)] the c.d.f $F$ of $X$ and $G$ of $Y$ are strictly increasing and differentiable;
	\item [ii)] the densities $g$ and $c$ are twice differentiable with continuous bounded second derivatives on their support. 
\end{enumerate}
Moreover, we assume that the kernels $K_0$ and $K$ satisfy the following: 
  \begin{enumerate}
	 \item [(i)] $K$ and $K_0$ are of bounded support and of bounded variation;
	 \item [(ii)] $0 \le K \le C$ and $0 \le K_0 \le C$ for some constant $C$;
	 \item [(iii)] $K$ and $K_0$ are first order kernels: $m_0(K)= 1$, $m_1(K)  = 0$ and $m_2(K) < +\infty$, and the same for $K_0$.
  \end{enumerate} 
In addition, in order to approximate $\hat c_n$ by $c_n$, we will  impose the slightly more stringent assumption on the bivariate kernel $K$, that it is twice differentiable with bounded second partial derivatives.

%\subsection{Heuristic}
%Recall that $\hat c_n(u,v)$ is the kernel copula density estimator made from the data $(F_n(X_i) , G_n(X_i)$, and $c_n(u,v)$ its analogue from the pseudo, but fixed with respect to $n$, data $( F(X_i) , G(X_i))$. The heuristic of the reason why our estimator works is that the $n^{-1/2}$ rate of convergence in uniform norm of $F_n$ and $G_n$  to $F$ and $G$   is faster than the  $1/\sqrt{na_n^2}$ rate of the non parametric kernel estimator $c_n$ of the copula density $c$. Therefore, the approximation step of the unknown transformations $F$ and $G$ by their empirical counterparts $F_n$ and $G_n$  does not have any impact asymptotically on the estimation step of $c$ by $c_n$.
% Put in another way, one can approximate $\hat c_n (F_n (x),G_n (y))$ by $c_n(F(x),G(y))$ at a faster rate than the convergence rate of $c_n(F(x),G(y))$ to $c(F(x),G(y))$. This is what is proved in the two approximation lemmas of section \ref{appendix}. The convergence properties of our estimator will then result from the well-known convergence properties of the kernel density estimators, which are also recalled in section \ref{appendix}.

\subsection{Weak  and strong consistency of the estimator}
We have the following pointwise weak consistency theorem:
\begin{theorem}\label{theorem3}
Let the regularity conditions on the densities and kernels be satisfied, if $h_n$ and $a_n$ tends to zero as $n \rightarrow \infty$ in such a way that $nh_n \rightarrow \infty$, $na_n^2 \rightarrow \infty$, then 
$$ \hat f_n (y|x) = f(y|x)+O_P\left({\frac{1}{{\sqrt {nh_n } }}+h_n^2+\frac{1}{\sqrt{na_n^2}}+a_n^2} \right).$$
\end{theorem}

\textbf{Proof.}
Recall from \ref{cn} and \ref{hatcn} that $c_n$ and $\hat c_n$ are estimators of the copula density $c$ based respectively on unobservable pseudo-data $(F(X_i),G(Y_i)$, and their approximations $(F_n(X_i),G_n(Y_i))$. The main ingredient of the proof follows from the decomposition:
\begin{align*}
 \hat f_n (y|x) - f(y|x) &= \hat g_n (y)\hat c_n (F_n (x),G_n (y)) - g(y)c(F(x),G(y)) \\ 
  &= \left[ {\hat g_n (y) - g(y)} \right]\hat c_n (F_n (x),G_n (y)) \\
  &+ g(y)\left[ {\hat c_n (F_n (x),G_n (y)) - c(F(x),G(y))} \right] \\ 
  :&= D_1  + D_2   
 \end{align*}
We proceed one step further in the decomposition of each terms, by first centering  at  fixed locations,
\begin{align}
 D_1  &= \left[ {\hat g_n (y) - g(y)} \right]\left[ {\hat c_n (F_n (x),G_n (y)) - \hat c_n (F(x),G(y))} \right] \nonumber\\ 
  &+ \left[ {\hat g_n (y) - g(y)} \right]\left[ {\hat c_n (F(x),G(y)) - c_n (F(x),G(y))} \right] \nonumber\\ 
  &+ \left[ {\hat g_n (y) - g(y)} \right]\left[ {c_n (F(x),G(y)) - c(F(x),G(y))} \right] \nonumber\\ 
  &+ \left[ {\hat g_n (y) - g(y)} \right]\left[ {c(F(x),G(y))} \right] \label{D1} \\
\nonumber  \\
 D_2  &= g(y)\left[ {\hat c_n (F_n (x),G_n (y)) - \hat c_n (F(x),G(y))} \right] \nonumber\\ 
  &+ g(y)\left[ {\hat c_n (F(x),G(y)) - c_n (F(x),G(y))} \right]\nonumber \\ 
  &+ g(y)\left[ {c_n (F(x),G(y)) - c(F(x),G(y))} \right] \label{D2}
 \end{align}
Convergence results for the kernel density estimators of section \ref{convdens1} entail that 
\begin{align*}
\hat g_n (y) - g(y) &=O_p(h_n^2+1/\sqrt{nh_n}) \\
c_n (F(x),G(y)) - c(F(x),G(y)) &=O_p(a_n^2+1/\sqrt{na_n^2})
\end{align*}
by lemma \ref{dens1} and \ref{dens2} respectively. 
Approximation lemmas \ref{approx1} and \ref{approx2} of sections \ref{sectionapprox1} and \ref{sectionapprox2} entail that 
\begin{align*}
\hat c_n (F(x),G(y))-c_n(F(x),G(y)) &=o_P ( a_n^2+1/\sqrt {na_n^2 })\\
\hat c_n (F_n (x),G_n (y)) - \hat c_n (F(x),G(y))&=  o_P (a_n^2+ 1/\sqrt {na_n^2 }).
\end{align*}
We therefore obtain that
 \begin{align*}
 D_1  &=  O_P \left( {h_n^2  + 1/\sqrt {nh_n } } \right)O_P \left( {a_n^2  + 1/\sqrt {na_n^2 } } \right) + O_P \left( {h_n^2  + 1/\sqrt {nh_n } } \right) \\ 
 D_2  & = o_P \left( {a_n^2+1/\sqrt {na_n^2 } } \right)  + O_P \left( {a_n^2  + 1/\sqrt {na_n^2 } } \right)
 \end{align*}
and the condition $a_n \rightarrow 0$, $h_n \rightarrow 0$, $na_n^2 \rightarrow +\infty$, $nh_n \rightarrow +\infty$
entails the convergence of the estimator.
\qed

\begin{remark}
As a corollary, we get the rate of convergence, by choosing the bandwidths which balance the bias and variance trade-off:
for an optimal choice of $h_n \simeq n^{-1/5}$ and $a_n  \simeq n^{-1/6}$, we get 
$$\hat f(y|x)=f(y|x)+O_P(n^{-1/3}).$$
Therefore, our estimator is rate optimal in the sense that it reaches the minimax rate $n^{-1/3}$ of convergence, according to Stone \cite{S1980}.
\end{remark}

% A MODIFIER
%We also have weak consistency results uniformly on sets: 
%\begin{corollary]
%If, in addition to the previous assumptions,  $g$ and $c$ also satisfies  assumption \textbf{(f'-0)},   $nh_n/\ln n \rightarrow \infty$ and  $na_n^2/\ln n \rightarrow \infty$, then,  for $x$ in the interior of $\mbox{supp}(f)$ and $[a,b]$ included in the interior of $\mbox{supp}(g)$, 
%$$ \mathop {\sup }\limits_{y \in [a,b]} |\hat f_n (y|x) - f(y|x)|=O_p\left((\ln n/na_n^2)^{1/2}\right).$$
%and,
%$$\mathop {\sup }\limits_{y \in R} \left| {\hat f_n (y|x) - f(y|x)} \right|
%=O_p\left(\left( {\frac{{\ln n}}{n}} \right)^{  1/3}  \right)$$
%\end{corollary]
%\textbf{Proof.}
%Use the same decomposition \ref{D1} and \ref{D2} as before and majorize in uniform norm. Then use the results in uniform norm of section \ref{appendix} in lemmas \ref{dens1} and \ref{dens2}.
%\qed

%\begin{remark}
%With the usual rates $h_n \simeq (\ln n/n)^{1/5}$, $a_n \simeq (\ln n/n)^{1/6}$, we get the rate
%$$ \mathop {\sup }\limits_{y \in [a,b]} |\hat f_n (y|x) - f(y|x)|=O_p\left((\ln n/n)^{1/3}\right)$$
%which is the optimal rate according to \cite{H1978}.
%\end{remark}
%-----------------------------

Almost sure results can be proved in the same way: we have the following  strong consistency result
\begin{theorem}
Let the regularity conditions on the densities and kernels be satisfied. If in addition $nh_n/(\ln \ln n) \to \infty$  and $na_n^2/(\ln \ln n) \to \infty$ , then
$$ \hat f_n (y|x) = f(y|x)+O_{a.s.}\left(a_n^2+\sqrt{\frac{\ln \ln n}{na_n^2}}+h_n^2+\sqrt{\frac{\ln \ln n}{nh_n}}\right).$$
\end{theorem}
\textbf{Proof.}
It follows the same lines as the preceding theorem, but uses the a.s. results of the consistency  of the kernel density estimators of lemmas \ref{dens1} and \ref{dens2}   and of the approximation lemmas \ref{approx1} and \ref{approx2}. It is therefore omitted.
\qed

\begin{remark}
For $h_n \simeq (\ln \ln n/n)^{1/5}$ and $a_n \simeq (\ln \ln n/n)^{1/6}$ which is the optimal trade-off between the bias and the stochastic term, one gets the optimal rate $(\ln \ln n/n)^{1/3}$.
\end{remark}

%------------------------------------------------------------------------
\subsection{Convergence in distribution}

\begin{theorem} \label{asympnormality}
%If  $nh_n \to \infty$, $na_n^2 \to \infty$, then
Let the regularity conditions on the densities and kernels be satisfied. $h_n\to 0$, $a_n\to 0$, $nh_n\to \infty$ and $na_n^2\to \infty$ entail
$$\sqrt{na_n^2}\left(\hat f_n (y|x) - f(y|x) \right) \stackrel{d}{\leadsto} \mathcal{N}\left(0,g(y)f(y|x)||K||_2^2 \right).$$
For $h_n\simeq n^{-1/5}$, $a_n  \simeq n^{-1/6}$ one gets the usual  rate $n^{-1/3}$.
\end{theorem}

\textbf{Proof.}
With the conditions on the bandwidths, all the terms in the previous decomposition \ref{D1} and \ref{D2},   are  negligible compared to $(na_n^2)^{-1/2}$ except $c_n (F(x),G(y)) - c(F(x),G(y))$, which is asymptotically normal  by the result of section \ref{appendix}, lemma \ref{dens2}
 $$\sqrt{na_n^2}g(y) \left[ {c_n (F(x),G(y)) - c(F(x),G(y))} \right]  \stackrel{d}{\leadsto} \mathcal{N}\left( {0,g^2(y)c(F(x),G(y))\left\| K \right\|_2^2  } \right).$$
An application of Slutsky's lemma yields the desired result.
\qed

For a vector $(y_1,\ldots,y_d)$, one can get a multidimensional version of the convergence in distribution (fidi convergence):
\begin{corollary}
With the same assumptions, for $(y_1,\ldots,y_d)$ in the interior of $\mbox{supp}(g)$ such that $g(y_i)f(y_i |x)\neq0$, \[
\sqrt{na_n^2} \left( {\left( {\frac{{\hat f_n (y_i |x) - f(y_i |x)}}{{\sqrt {g(y_i)f(y_i |x)} \left\| K \right\|_2 }}} \right),i = 1,...,m} \right) \stackrel{d}{\leadsto} N^{(m)} 
\]
 where $N^{(m)}$ is the standard $m$-variate centered normal distribution with identity variance matrix.
\end{corollary}
\textbf{Proof.} It simply follows from the use of the Cramér-Wold device and is therefore omitted. For details, see e.g. \cite{B1998}, theorem 2.3.
\qed

%-----------------------------------------------------------------
\subsection{Asymptotic Bias, Variance and Mean square error}

The asymptotic bias is calculated in the following proposition.
\begin{proposition}
With the assumptions of Theorem \ref{theorem3}, we have
\begin{align*}
B_0:=E(\hat f_n (y|x)) - f(y|x) 
= g(y) B_K(c,x,y)\frac{{a_n^{2} }}
{2} + o(a_n^2)
\end {align*}
with 
$ B_K(c,x,y):=m_2 (K_1 )\frac{{\partial ^2 c(F(x),G(y))}}{{\partial u^2 }}
+
m_2 (K_2 )\frac{{\partial ^2 c(F(x),G(y))}}{{\partial v^2 }}
$.
\end{proposition}

\textbf{Proof.}
(Sketch). By taking expectation in the decomposition \ref{D1} and \ref{D2}, 
\begin{align*}
 ED_1 &=c(F(x),G(y))E[\hat g_n(y) -g(y)]+R_1\\
 ED_2 &=g(y)E\left([c_n(F(x),G(y))-c(F(x),G(y))]\right)+R_2
 \end{align*}
where we made appear the bias of $\hat g_n$ and $c_n$ and where $R_1$ and $R_2$ stand for the remaining terms. With the assumptions on the bandwidths and derivations made tedious by the transformation of the data by the empirical margins, (see Fermanian \cite{F2005} theorem 1 for such a calculation), the terms in $R_2$ are negligible compared to the bias of $c_n$. The bias of $c_n$, which is simply the bias of a bivariate kernel density estimator, is of order $a_n^2$.  Similarly, by  bounding the product terms in $D_1$ by Cauchy-Schwarz inequality, routine  analysis show that the terms in $R_1$ are negligible compared to the bias of $\hat g_n$, which is of order $h_n^2$. Since $h_n^2$ is itself negligible to $a_n^2$, the main term in the decomposition is $g(y)E(c_n(F(x),G(y))-C(F(x),G(y)))$. Plugging the expression of the bias given in lemma \ref{dens2}, yields the desired result. 
\qed

The asymptotic variance has already been derived in theorem \ref{asympnormality},
 $$V_0:=Var(\hat f(y|x))=1/(na_n^2)g(y)f(y|x)||K||_2^2+o(1/(na_n^2)).$$ 
Together with the computation of the asymptotic bias, we get  the asymptotic mean squared error as a corollary:
\begin{corollary} With the previous assumptions, the Asymptotic Mean Squared Error (AMSE) $E_0$ at $(x,y)$ is 
\begin{align*}
	E_0 &:=B_0^2+V_0\\
	&= \frac{{a_n^4g^2(y)\left( { B_k(c,x,y)} \right)^2 }}
{4} + \frac{g(y)f(y|x)||K||_2^2}{na_n^2}  + o\left(a_n^4+\frac {1}{na_n^2}\right)
\end{align*}
which gives, for the choice of the usual bandwidths mentioned above,
\[
E_0=  n^{-2/3} g^2 (y)\left( {\frac{{B_K^2(c,x,y) }}
{4} + c(F(x),G(y))||K||_2^2 } \right) + o(n^{ - 2/3} ).
\]
\end{corollary}

%%Bosq and Bleuez \cite{BB1978} and Scott and Terell \cite{S1992} theorem 6.6 have shown the generality of the kernel method, in the sense that the only square integrable unbiased estimator of the density for the former authors, or that  every  density estimator which is a smooth functional  of the empirical distribution function  for the latter, can be  written as  kernel estimators. Therefore, we conjecture that our approach can be extended to other nonparametric estimators like projection or local polynomial ones.
%\end{conj}

%---------------------------------------------------------------------------------------------------------
%---------------------------------------------------------------------------------------------------------
\section{Comparison with other estimators}
\subsection{Presentation of alternative estimators}
For convenience, we recall below the definition of other estimators of the conditional density encountered in the literature and summarize their bias and variance properties. We will note the bias of the ith estimator  $\hat f_n^i(y|x)$ by $E_i$ and its variance by $V_i$.
\begin{enumerate}
	\item \textbf{Double kernel estimator}: as defined in the introduction section of our paper by the following ratio,
\[
\hat f_n^{(1)} (y|x) := \frac{{\frac{1}{n}\sum\limits_{i = 1}^n {K'_{h_1} (X_i  - x)K_{h_2} (Y_i  - y)} }}{{\frac{1}{n}\sum\limits_{i = 1}^n {K'_{h_1} (X_i  - x)} }}.
\]
  where $h_1$ and $h_2$ are the bandwidths. One then have, see e.g. \cite{HBG1996},
   \begin{itemize}
  	\item Bias:
\begin{align*}
 B_1 &= \frac{{h_1^2 m_2(K) }}{2}\left( {2\frac{{f'(x)}}{{f(x)}}\frac{{\partial f(y|x)}}{{\partial x}} +\frac{{\partial ^2 f(y|x)}}{{\partial x^2 }}+\left( {\frac{h_2}{{h_1}}} \right)^2 \frac{{\partial ^2 f(y|x)}}{{\partial y^2 }}} \right) \\ 
 &+  o\left( {h_1^2+h_2^2  } \right)  
 \end{align*}
	  \item Variance: \[
V_1 = \frac{{\left\| K \right\|_2^2 f(y|x)}}{{nh_1h_2f(x)}}\left( {\left\| K \right\|_2^2  - h_2f(y|x)} \right)  +o\left( {\frac{1}
{{nh_1 h_2 }}} \right)
\]
   \end{itemize}
\item \textbf{Local polynomial estimator}:
Set \[
R(\theta ,x,y): = \sum\limits_{i = 1}^n {\left( {K_{h_2} (Y_i  - y) - \sum\nolimits_{j = 0}^r {\theta _j (X_i  - x)^j } } \right)^2 K'_{h_1} (X_i  - x)} ,
\]
then the local polynomial estimator is defined as\[
\hat f_n^{(2)} (y|x): = \hat \theta _0 ,
\]
where $\hat \theta_{xy}:=(\hat \theta _0 ,\hat \theta _1 , \ldots ,\hat \theta _r )$ is  the value of $\theta$ which minimizes $R(\theta ,x,y)$.
This local polynomial estimator, although it has a superior bias than the kernel one, is no longer restricted to be non-negative and does not integrate to 1, except in the special case $r=0$.
From results of \cite{FYT1996}, we get for the local linear estimator (see also \cite{FY2005} p. 256),
  \begin{itemize}
	\item Bias: \[
B_2= \frac{{h_1^2 m_2(K') }}{2}\frac{{\partial ^2 f(y|x)}}{{\partial x^2 }} + \frac{{h_2^2 m_2(K) }}{2}\frac{{\partial ^2 f(y|x)}}{{\partial y^2 }} + o(h_1^2  + h_2^2 )
\]
	\item Variance: \[
V_2 = \frac{{||K||_2^2 ||K'||_2^2 f(y|x)}}{{nh_1 h_2 f(x)}}+o\left( {\frac{1}
{{nh_1 h_2 }}} \right)
\]
  \end{itemize}
\item \textbf{Local parametric estimator}: As in \cite{HY2002} and \cite{FY2005}, set
\[
R_1 (\theta ,x,y): = \sum\limits_{i = 1}^n {\left( {K_{h_2} (Y_i  - y) - A(X_i  - x,\theta )} \right)^2 K'_{h_1} (X_i  - x)} 
\]
where
$
A(x,\theta ) = l\left( {\sum\nolimits_{j = 0}^r {\theta _j (X_i  - x)^j } } \right)
$
and $l(.)$ is a monotonic function mapping $\mathds{R} \mapsto \mathds{R}^+$, e.g. $l(u)=\exp(u)$.
Then, \[
\hat f_n^{(3)} (y|x): = A(0,\hat \theta ) = l(\hat \theta _0 ).
\]
  \begin{itemize}
	\item Bias:
\begin{align*}
	B_3  &= h_1^2 \eta (K')\left( {\frac{{\partial ^2 f(y|x)}}
{{\partial x^2 }} - \frac{{\partial ^2 A(0,\theta _{xy} )}}
{{\partial x^2 }}} \right) + \frac{{h_2^2 m_2 (K)}}
{2}\frac{{\partial ^2 f(y|x)}}
{{\partial y^2 }} \\
&+ o(h_1^2  + h_2^2 )
\end{align*}
  \item Variance:\[
V_3 = \frac{{\tau (K,K')^2 f(y|x)}}{{nh_1h_2f(x)}}+o\left( {\frac{1}{{nh_1h_2 }}} \right)
\]
  \end{itemize}
where $\eta$ and $\tau$ are kernel dependent constants.

\item \textbf{Constrained local polynomial estimator}:
A simple device to force the local polynomial estimator to be positive is to set $\theta _0  = \exp (\alpha )$ in the definition of $R_0$ to be minimized. The constrained local polynomial estimator $\hat f_n^4(y|x)$ is then defined analogously as the local polynomial estimator $\hat f_n^2(y|x)$.
We have, as in  \cite{HY2002} and \cite{FY2005}:
%\[
%\hat f_n^{(4)} (y|x) - f(y|x) = \left( {nhb} \right)^{ - 1/2} \left( {\frac{{\left\| K \right\|_2^2 f(y|x)}}{{f(x)}}} %\right)^{1/2} N_{n2}  + h^2 \frac{{\kappa _2 }}{2}f^{(2)} (y|x) + b\frac{{\mu _2 }}{2}f^{(2)} (y|x) + o\left( {(nhb)^{ %- 1/2}  + h^2  + b^2 } \right)
%\]
\begin{itemize}
 \item Bias: \[
B_4 : = h_1^2 \frac{{m_2 (K')}}
{2}\frac{{\partial ^2 f(y|x)}}
{{\partial x^2 }} + h_2^2 \frac{{m_2 (K)}}
{2}\frac{{\partial ^2 f(y|x)}}
{{\partial y^2 }} + o(h_1^2  + h_2^2 )
\]
 \item Variance: $$V_4 = \frac{{\left\| K \right\|_2^2 f(y|x)}}{{nh_1h_2f(x)}}+o\left( {\frac{1}{{nh_1h_2 }}} \right)$$
 \end{itemize}
\end{enumerate}

\subsection{Asymptotic Bias and Variance comparison}

All estimators have (hopefully) the same order $n^{-1/3}$ and $n^{-2/3}$ in their asymptotic bias and variance terms, for the usual bandwidths choice. The main difference lies in the constant terms which depend on unknown densities. 

\textbf{Bias}:
Contrary to all the alternative estimators whose bias involves derivatives of the full conditional density,  one can note that our estimator's bias only involves the density of $Y$ and the derivatives of the copula density. To make things more explicit, the  terms involved, e.g. in the local polynomial estimator, write themselves as the sum of the derivatives of the conditional density,
\begin{align*}
  h_n^{-2} B_2 \approx  \frac{{\partial ^2 f(y|x)}}
{{\partial x^2 }} + \frac{{\partial ^2 f(y|x)}}
{{\partial y^2 }}
\end{align*}
that is to say,
\begin{align*}
h_n^{-2} B_2&\approx f'(x)g(y)\frac{{\partial c(F(x),G(y))}}
{{\partial u}} + f^2 (x)g(y)\frac{{\partial ^2 c(F(x),G(y))}}
{{\partial u^2 }}  \\
   &+ 2g'(y)g(y)\frac{{\partial c(F(x),G(y))}}
{{\partial v}} + g^3 (y)\frac{{\partial ^2 c(F(x),G(y))}}
{{\partial v^2 }}  
\end{align*}
whereas our $(g(y)/2)B_K( c,x,y)$ term, modulo the constants involved by the kernel,  is written as
\begin{align*}
 a_n^{-2} B_0\approx g(y)\left( {\frac{{\partial ^2 c(F(x),G(y))}}
{{\partial u^2 }} + \frac{{\partial ^2 c(F(x),G(y))}}
{{\partial v^2 }}  } \right).
\end{align*}
It then becomes clear that we have a simpler expression, with less unknown terms, as is the case for competitors which do involve the density $f$ and its derivative $f'$ of $X$  and the derivative $g'$ of the $Y$ density. 

In a fixed bandwidth and asymptotic context, it seems difficult to compare further. Nonetheless, we believe  this feature of our estimator would be practically relevant when it comes to choosing the bandwidths. Indeed, bandwidth selection is usually performed by minimizing local or global  asymptotic error criteria such as Asymptotic Mean Square Error (AMSE) or Asymptotic Mean Integrated Square Error (AMISE), in which unknown terms have to be estimated. Since in our approach, the asymptotic bias and variance involve less unknown terms, we expect that a higher accuracy could be obtained in this pre-estimation stage. Moreover, by having managed to separate the estimation problem of the marginal from the copula density, we could use known optimal data-dependent bandwidths selection procedures for density estimation such as cross validation, separately for the density of $Y$ and for the copula density. 
%Such a study in the finite sample setting will be performed in a forthcoming paper. 

\begin{remark}
Since the copula density $c$ has a compact support $[0,1]^2$, our estimator may suffer from bias issues on the boundaries, i.e. in the tails of $X$ and $Y$. To correct these issues, one could apply one of the several known techniques  to reduce the bias of the kernel estimator on the edges (see e.g \cite{FY2005} chapter 5.5, boundary kernels, reflection, transformation and  local polynomial fitting). In the tail of the distribution of $X$, this bias issue in the copula density estimator is balanced by the improved variance, as shown below.
\end{remark}

\textbf{Variance}:
The variance of our estimator involves a product of the density $g(y)$ of $Y$ by the conditional density  $f(y|x)$,
$$na_n^2V_0 \approx g(y)f(y|x)=g^2(y)c(F(x),G(y)$$ whereas  competitors involve the ratio of $f(y|x)$ by the density $f(x)$ of $X$
$$\frac{f(y|x)}{f(x)}=\frac{g(y)}{f(x)}c(F(x),G(y)).$$ It is a remarkable feature of the estimator we propose, that its variance does not involve directly $f(x)$, as is the case for the competitors, but only its contribution to $Y$, through the copula density. This reflects the ability announced in the introduction of the copula representation to have effectively separated the randomness pertaining to $Y$ alone, from the dependence structure of $(X,Y)$.  Moreover, our estimator also does not suffer from the unstable nature of competitors who, due to their intrinsic ratio structure, get an explosive  variance for small value of the density $f(x)$, making conditional estimation difficult, e.g. in the tail of the distribution of $X$. 
%As for the bias, a comparison with simulations for finite samples will be done in a forthcoming paper.

\begin{remark}
To make estimators comparable, we have restricted ourselves to so-called fixed bandwidths estimators, i.e. nonparametric estimators where the bandwidths are of the generic form $h_n=bn^{\alpha}$ or $h_n=b(\ln n/n)^\alpha$ with $\alpha$ and $b$ real numbers. Improved behavior for all the preceding estimators can be obtained with data-dependent bandwidths where $h_n=H_n(X_1,\ldots,X_n,x)$ can be functions of the location and of the data. 
\end{remark}

\subsection{Finite sample numerical simulation}

\subsubsection{Practical implementation of the estimator}
Although the proposed estimator seems to compare favorably asymptotically, some pitfalls linked to the copula density estimation may show up in the practical implementation:

\textbf{Infinities at the corners:} many copula densities exhibit infinite values at their corners. Therefore, to avoid that $(F_n(X_{i}),G_n(Y_{i}))$ be equal to $(1,1)$, we change the empirical distribution functions $F_n$ and $G_n$ to $n/(n+1)F_n$ and $n/(n+1)G_n$ respectively.

 \textbf{Boundary bias:} since the copula density is of compact support $[0,1]^2$, the kernel method of estimation may suffer from boundary bias. To alleviate this issue, we suggest to use boundary-corrected kernels such as the beta kernels $K_{x,b}(t)=\beta_{x/b+1,(1-x)/b+1}(t)$, where $\beta_{a,b}(t)$ denotes the pdf of a Beta(a,b) distribution, advocated by Chen \cite{C1999}, and used e.g. by \cite{GHNS2007} for estimating  loss distributions. The modified  copula density pseudo estimator is thus defined as  
$c_n(u,v)=n^{-1}\sum_{i=1}^n {K_{u,a_n}(U_i)K_{v,a_n}(V_i)}$.

 \textbf{Bandwidth selection:} performance of nonparametric estimators depends crucially on the bandwidths. For conditional density, bandwidth selection is a more delicate matter than for density estimation due to the  multidimensional nature of the problem. Moreover, for ratio-type estimators, the difficulty is increased by the local dependence of the bandwidths $h_y$ on $h_x$ implied by conditioning near $x$. For the copula estimator, a supplemental issue comes from the fact that the pseudo-data $F(X_i),G(Y_i)$ is not directly accessible. 
Inspection of the AMISE of the copula-based estimator suggest we can separate the bandwidth choice of $h$ for $\hat g(y)$ from the bandwidth choice of $a_n$ the copula density estimator $\hat c_n$.
A rationale for a  data-dependent method is to separately select $h$  on the $Y_i$  data alone (e.g. by cross-validation or plug-in), from the  $a_n$ of the copula density $c$ based on the approximate data  $F_n(X_i),G_n(Y_i)$. However, such a bandwidth selection would require deeper analysis and we leave  a detailed study of a practical data-dependent method for bandwidth selection of the copula-quantile estimator, together with a global and local comparison of the estimators at their respective optimal bandwidths  for further research.

\subsubsection{Model and comparison results}
We simulated a sample of $n=100$  variables $(X_i,Y_i)$, from the following model:
%\textbf{Model 1:}
$X,Y$ is marginally distributed as $\mathcal{N}(0,1)$   and  linked via Frank Copula . $$C(u,v,\theta)=\frac{\ln[(\theta+\theta^{u+v}-\theta^u-\theta^v)/(\theta-1)]}{\ln \theta}  $$ with parameter $\theta=100$.

%\textbf{Model 2:}
%We study the nonlinear regression model studied by \cite{FYT1996},
%$Y=0.23X(16-X)+0.4\epsilon$
%with $\epsilon\sim\mathcal{N}(0,1)$ is independent of $X\sim U_{[5,11]}$. 

\begin{figure}
\centering
\includegraphics[width=13cm]{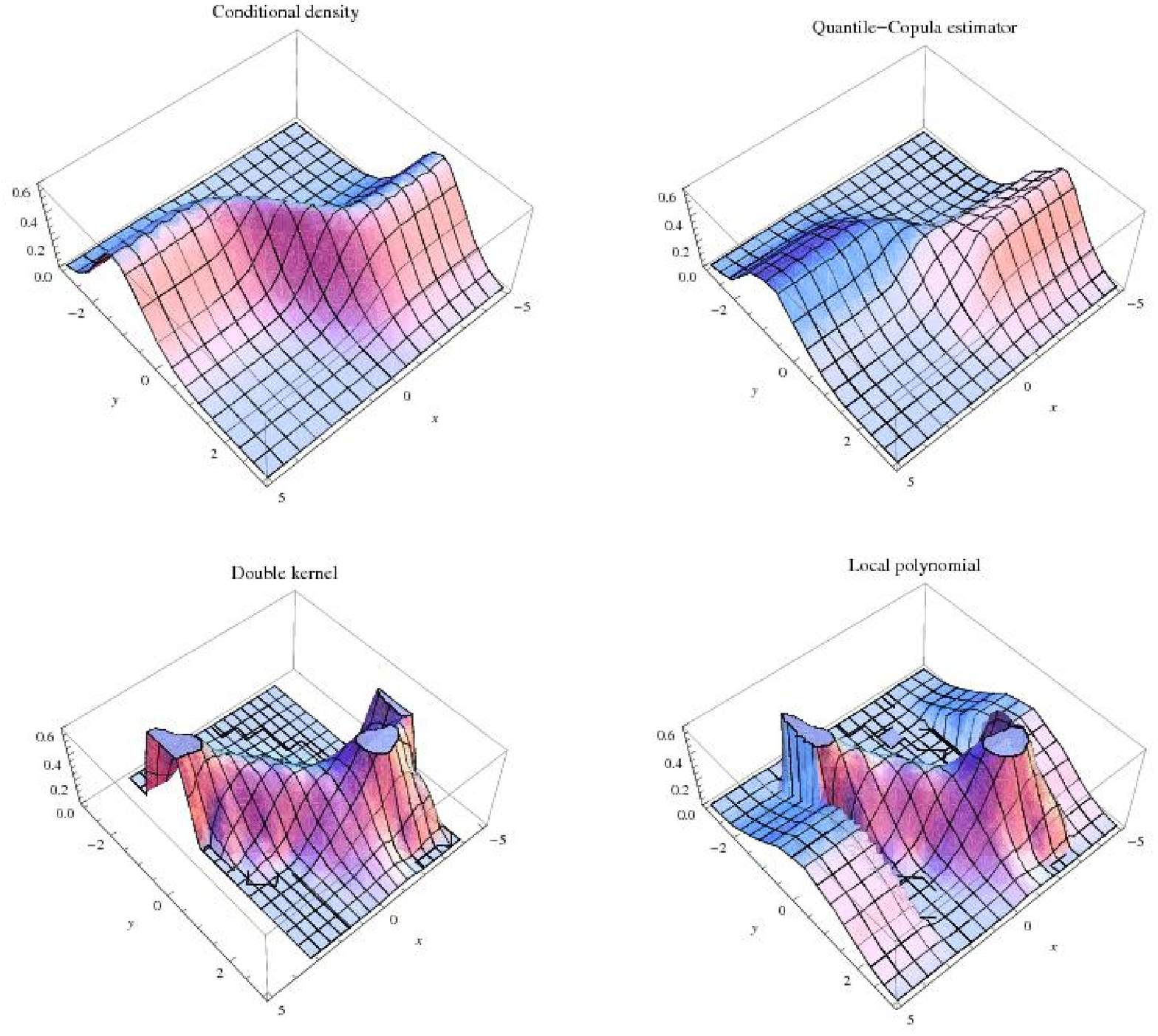}
\caption[]{3D Plots. From left to right, top to bottom: true density, quantile-copula estimator, double kernel, local polynomial (clipped).}
\label{fig1}
%\end{figure}

%\begin{figure}[h]
%\centering
\includegraphics[width=13cm]{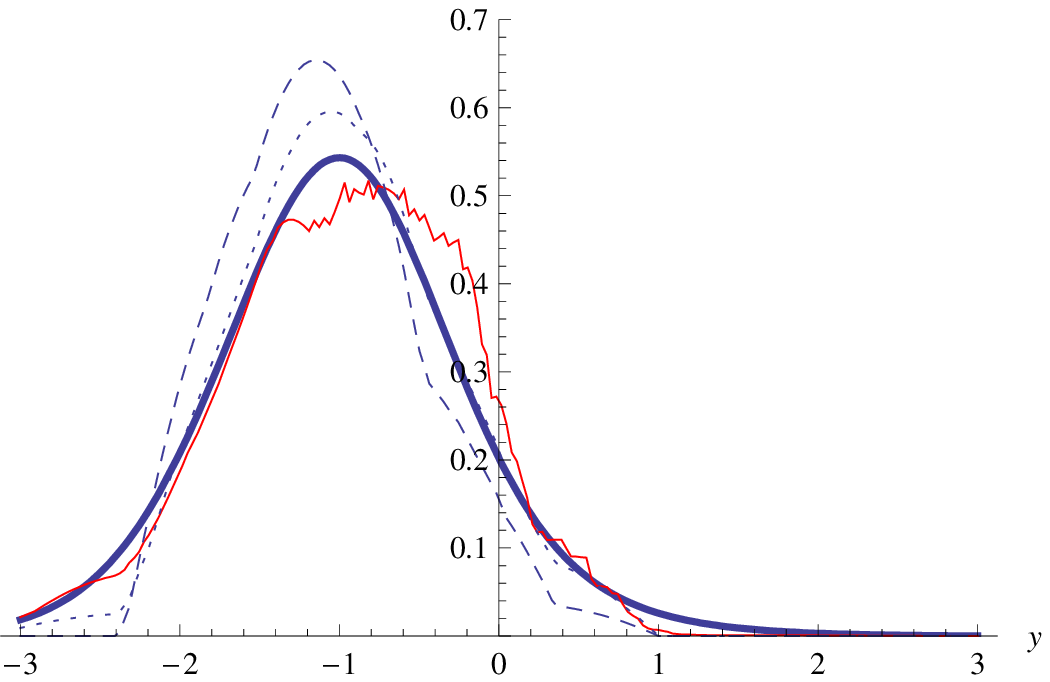}
\caption{Comparison at x=2: conditional density=thick curve, quantile-copula=continuous line, double kernel=dotted curve, local polynomial=dashed curve.}
\label{fig2}
\end{figure}

We restricted ourselves to simple, fixed for all $x,y$, rule-of-thumb methods based on Normal reference rule to get a first picture.  For  the selection of $a_n$ of the copula density estimator, we applied Scott's Rule on the data $F_n(X_i)$. We used Epanechnikov kernels for $\hat g(y)$ and the other estimators. We plotted the conditional density along with its estimations on the domain $x\in[-5,5]$ and $y\in[-3,3]$  on  figure \ref{fig1}. A comparison plot at $x=2$ is shown on figure \ref{fig2}. % and for model 2 on the domain $x\in[5,11]$ and $y\in[11,16]$ on figure \ref{fig2}. 

%\begin{figure}
%\includegraphics[width=14cm]{fig2}
%\caption{3D Plots for regression model 2}
%\label{fig2}
%\end{figure}

%Some strategies proposed in the literature for the alternative estimators include the rule-of-thumb methods based on the assumption that the conditional density is Normal with a linear regression by \cite{BH2001}, the two step  mixed rule by \cite{FYT1996} who combine rule-of-thumb and cross-validation, the least-squares cross validation by \cite{FY2004} and \cite{HRL2004}, and the bootstrap approach by \cite{HWY1999}.

%Note that competitors change the bandwidths for each $x$.
%In order to make a fair comparison, we have to compare the estimators irrespectively of the bandwidth.

\subsubsection{Clipping and Estimation in the tails}
As mentioned earlier, as the performance of the estimators depends on the performance of the bandwidths selection method, it is delicate to give a conclusive answer. However, we would like to illustrate at least one case where the proposed estimator clearly outperforms its competitors. Indeed, one major issue of alternative estimators already mentioned is their numerical explosion when the estimated density $\hat f(x)$ is close to zero. In particular, if the kernel is of compact support, the denominator is zero for the $x$ whose distance from the closest $X_i$ exceeds half  the bandwidth times the length of the support, thereby allowing estimation only on a closed subset of $X$ included in $[\min X_i,\max X_i]$.   This is one of the reason why simulation studies are often performed either with a marginal $X$ density of bounded support and/or with a Gaussian kernel. Note that the problem remains with a Gaussian kernel since the estimated density can become quickly lower than the machine precision.
To prevent from this numerical explosion, the definition of the conditional density estimators have to be modified either by  
$$
\hat f(y|x) = \left\{ {\begin{array}{*{20}c}
   {\frac{{\hat f_{XY} (x,y)}}
{{\hat f_X (x)}}} & {\mbox{if }\hat f_X (x) > c}  \\
   {\hat a(y)} & {\mbox{if }\hat f_X (x) = 0}  \\
 \end{array} } \right.
 \mbox{  or by,  } 
\hat f(y|x) =\frac{{\hat f_{XY} (x,y)}}
{{\max\{\hat f(x),c\}}}$$
where $c>0$ is an arbitrary amount of clipping, and  $\hat a(.)$ is an arbitrary density estimator (usually chosen to be zero or $\hat g(y)$).

An illustration of these issues clearly appears in figure \ref{fig1}. The unclipped version of the double kernel estimator is unable to estimate the conditional density for $|x|$ roughly $>3$, and the clipped version of the local polynomial estimator with $c=0.00001$ and $\hat a(y)=\hat g(y)$ gives a wrong estimation in the tails, reflecting the arbitrary choices in the clipping decision.  To the contrary, the quantile-copula estimator is surprisingly  able to estimate the conditional density $f(y|x)$ at locations $x$ where there is ``no data'', i.e. in the tails of the distribution of $X$. 
An explanation of this apparently paradoxal phenomenon comes from the fact that the estimator is partially based on the ranks of $X_i$ and $Y_i$. Therefore, it can recover ``hidden'' information on the density of $X$ from the ordering of the pairs $(X_i,Y_i)$. See Hoff \cite{H2007} for a detailed explanation. We believe that this feature might be of potential interest for applications, e.g. in statistical inference of extreme values and rare events.

%We believe an explanation of these results is connected with the deep probabilistic issues associated with the definition of conditional probability with a conditioning event of probability zero.
%Indeed, the usual way to define the conditional density by
%\[
%f(y|x) = \left\{ {\begin{array}{*{20}c}
%   {\frac{{f_{XY} (x,y)}}
%{{f_X (x)}}} & {\mbox{if }f_X (x) > 0}  \\
%   {a(y)} & {\mbox{if } f_X (x) = 0}  \\
% \end{array} } \right.
%\]
%where $a(.)$ is an arbitrary density function, usually chosen to be $a=0$, leads to different values of conditional probabilities, as exemplified e.g. by the Borel-Kolmogorov or Kac-Slepian paradoxes (See Rao [2005] Conditional Measures and Applications, 2nd Edition, Springer).

%---------------------------------------------------------------------------------------------------------
%---------------------------------------------------------------------------------------------------------
\section*{Discussion}
The quantile transform and use of the copula formula has thus turned the conditional density formula \eqref{ratio} of the ratio type into the  product one \eqref{densitecondi}. This formula was the backbone of our article where this product form appeared to be especially appealing for statistical estimation: consistency and limit results where obtained by simple combination of the previous known ones on (unconditional) density estimation. %It allowed to bypass the well known fact that Probability does not mix well with Algebra, in the sense that the law of a random variable transformed by an algebraic operation, and especially the inverse, is not easy to tackle with. 
The estimator obtained shows interesting asymptotic bias and variances properties compared to competitors. Although its finite sample implementation does not give yet a clear and conclusive picture, it already yields some promising results, e.g. for estimation in the tails of $X$, where the proposed estimator does not suffer from clipping issues. 
%Moreover, we stress that the limit results obtained do not make use of the commonly stated hypothesis in regression that the marginal density $f$ be bounded below by a strictly positive constant on a compact subset.

%-----------------------------------------------------------------------------
%------------------------------------------------------------------------------

\section{Appendix : auxiliary results} \label{appendix}
In this section, we gather some preliminary results which we will need as basic tools for the demonstrations of section 3. In subsection \ref{approxglivenko}, we recall classical results about the convergence of the Kolmogorov-Sminorv statistic. Next, we make a brief overview of kernel density estimation  and apply these results to the estimators $\hat g_n$ (section \ref{convdens1}) and $c_n$ (section \ref{convdens2}). Eventually, we need two approximation lemmas of $\hat c_n$ by $c_n$ in sections \ref{sectionapprox1} and \ref{sectionapprox2}.

\subsection{Approximation of the pseudo-variables $F(X_i)$ by  their estimates $F_n(X_i)$} \label{approxglivenko}
  For $(X_i,i=1,\ldots,n)$  an i.i.d. sample of a real random variable $X$ with common c.d.f. $F$, the Kolmogorov-Smirnov  statistic is defined as $D_n:=\left\| {F_n  - F} \right\|_\infty$. Glivenko-Cantelli,  Kolmogorov and Smirnov, Chung, Donsker among others have studied its convergence properties in increasing generality (See \cite{SW1986} and \cite{VDVW1996} for recent accounts).  For our purpose, we only need to formulate these results in the following rough form:

\begin{lemma}\label{glivenko} 
For an i.i.d. sample from a continuous c.d.f. $F$,
\begin{align}
	 \left\| {F_n  - F} \right\|_\infty &=  O_{a.s.}\left(\sqrt {\frac{{\ln \ln n}}{n}} \right) \label{gli1} \\
   \left\| {F_n  - F} \right\|_\infty   &= O_P \left( {\frac{1}{{\sqrt n }}} \right) \label{gli2}.
\end{align}
\end{lemma}

Since $F$ is unknown, the random variables $U_i=F(X_i)$ are not observed. As a consequence of the preceding lemma \ref{glivenko}, one can naturally approximate these variables by the statistics $F_n(X_i)$. Indeed,
$$\left| {F(X_i ) - F_n (X_i )} \right| \le \mathop {\sup }\limits_{x \in R} \left| {F(x) - F_n (x)} \right|=\left\| {F_n  - F} \right\|_\infty \quad \mbox{a.s.}$$
Thus, $|F(X_i ) - F_n (X_i )|$ is no more than an $O_P((\ln\ln n/n)^{1/2})$ or an $O_{a.s.}(n^{-1/2})$. 
%\begin{align}
%	\left| } \right| &=  O_{a.s.}\left( {\sqrt {\frac{{\ln \ln n}}{n}} } \right) \label{approx F} \\
%  \left| {F(X_i ) - F_n (X_i )} \right| &= O_P \left( {\frac{1}{{\sqrt n }}} \right)
%\end{align}
These rates of approximation appears to be faster than those of statistical estimation of densities, as is shown in the next subsection.
%-----------------------------------------------

\subsection {Convergence of the kernel density estimator $\hat g_n$} \label{convdens1}
We recall below some classical results about the convergence of the Parzen-Rosenblatt kernel non-parametric estimator $\hat f_n$ of a d-variate density. Since its inception by Rosenblatt \cite{R1956} and Parzen \cite{P1962}, it has been studied by a great deal of authors.  See e.g. Scott \cite{S1992}, Prakasa Rao \cite{PR1983}, Nadaraya \cite{N1989} for details. See also Bosq \cite{B1998} chapter 2. 

It is well known that the bias of the kernel density estimator depends on the degree of smoothness of the underlying density, measured by its number of derivatives or its Lipschitz order. In order to get the convergence of the bias to zero, it suffices to assume that the density is continuous (See \cite{P1962}). To get further information on the rate of convergence of the estimator, it is necessary to make further assumptions.  Moreover, for kernel functions with unbounded support, the rate of convergence also depends on the tail behavior of the kernel (See Stute \cite{S1982}). Therefore, for clarity of exposition and simplicity of notations, we will make the customary assumptions that the density is twice differentiable and that the kernel is of bounded support. We then have the following results:
\begin{itemize}
	\item Bias:   With the previous assumptions,  for a $x$ in the interior of 
$supp(f)$,  $h_n \rightarrow 0$ and $nh_n^d \rightarrow \infty$ entail that
\[
E\hat f_n (x) = f(x) + 
\frac{{h_n^2 }}{2}\int\limits_{\mathds{R}^d } {\sum\limits_{1 \le i,j \le d} {\frac{{\partial ^2 f(x)}}{{\partial x_i \partial x_j }}z_i z_j K(z)dz} }  + o(h_n^2 ).
\]
With the multivariate kernel $K$ as a product of $d$ order one kernels $K_i$, the above sum reduces to the diagonal terms.
\[
E\hat f_n (x) = f(x) + 
\frac{{h_n^2 }}{2} \sum\limits_{1 \le i \le d}{m_2(K_i)\frac{{\partial ^2 f(x)}}{{\partial x_i^2 }}} + o(h_n^2 ).
\]
  \item Variance: with the same assumptions,  
\[
Var\left[ {\hat f_n (x)} \right] = \frac{{f(x)}}{{nh_n^d }}\left\| K \right\|_2^2  + o\left( {\frac{1}{{nh_n^d }}} \right).
\]
  \item Pointwise asymptotic normality: under the previous conditions, 
\[
\sqrt {nh_n^d } \left( {\hat f_n (x) - E \hat f_n(x) } \right) \stackrel{d}{\leadsto} \mathcal{N}(0,f(x)\left\| K \right\|_2^2 ). 
\]

For a choice of the bandwidth as $h_n \simeq n^{-1/(d+4)}$, which realizes the optimal trade-off between the bias and variance,   one gets the rate $n^{-2/(d+4)}$, which is the optimal speed of convergence in the minimax sense in the class of density functions with bounded second derivatives, according to \cite{S1980}.

%One can refine these results by a chaining argument to get uniform rate of convergence on a compact set (see Bickel and Rosenblatt \cite{BR1973}): for $f$ bounded and non-vanishing on a compact subset $J$ included in the interior of $supp(f)$,
%\[
%\mathop {\sup }\limits_{x \in J} \left| {\hat f_n (x) - E\hat f_n (x)} \right| = O_p \left[ {\left( {\frac{{\ln n}}{{nh_n }}} \right)^{1/2} } \right].
%\]
%Therefore, for the choice of the bandwidth $h_n  \simeq ( \ln n/n )^{1/d+4} $ which realizes the optimal trade-off between the bias and variance, one gets the following result in probability:
%\[
%\mathop {\sup }\limits_{x \in J} \left| {\hat f_n (x) - f(x)} \right| = O_p \left[ {\left( {\frac{{\ln n}}{n}} \right)^{2/(d+4)} } \right]
%\]
%which is the optimal speed in the minimax sense in the class of density functions with bounded second derivatives, according to Hasminskii \cite{H1978}.

\item Pointwise almost sure convergence: if moreover $nh_n^d/(\ln \ln n) \to \infty$ (see \cite{D1974}),  we have that  
$$\hat f_n (x) - E\hat f_n(x) = O_{a.s.} \left( {\sqrt {\frac{{\ln \ln n}}{{nh_n^d }}} } \right).
$$
For a choice of the bandwidth as $h_n \simeq \left((\ln\ln n)/n\right)^{1/(d+4)}$, we get the rate of convergence $\left((\ln \ln n)/n\right)^{2/(d+4)}$:
 $$
\hat f_n (x) - f(x) = O_{a.s.} \left( {\left(\frac{\ln \ln n}{n}\right)^{2/(d+4)}} \right).
$$%\item on $\mathds{R}$ , for every $k \in \mathbb{N}$, set $\log _1 n = \ln (n  \vee  e)$ and $\log _k n = \ln (\log _{k - 1} n)$. Then, 
%\[
%\left\| {\hat f_n (y) - f(y)} \right\|_\infty   = O_{a.s.} \left( {\frac{{\log _k n}}
%{1}\left( {\frac{{\ln n}}
%{n}} \right)^{2/5} } \right)
%\]
\end{itemize}

Applied to our case ($d=1$), we can summarize these results for further reference in the following lemma for the estimator $\hat g_n $ of the density $g$ of $Y$:
\begin{lemma} \label{dens1}
 With the previous assumptions, for a point $y$ in the interior of the support of $g$, and a bandwidth chosen such as $h_n  \simeq n ^{-1/5} $, we have 
\begin{align*}
\left| {\hat g_n (y) - g(y)} \right| &= O_p (n^{-2/5}) \\
n^{2/5} \left[ {\hat g_n (y) - g(y)} \right]  &\stackrel{d}{\leadsto} \mathcal{N}\left( {0,g(y)\left\| K_0 \right\|_2^2 } \right).
\end{align*}
With the same assumptions, but for a bandwidth choice of $h_n  \simeq ( \ln\ln n/n )^{1/5} $, 
\begin{align}
\hat g_n (y) - g(y) = O_{a.s.} \left( {\left( {\frac{{\ln \ln n}}
{n}} \right)^{2/5} } \right).
\end{align}
\end{lemma}

%\noindent
%\textbf{Assumption $H_1$ :} The bandwidth is chosen such that $h_n  = O(( \ln n/n )^{1/5} )$

%-----------------------------------------------------------------------------------
\subsection {Convergence of $c_n(u,v)$} \label{convdens2}

As mentioned before, the assumptions that $F$ and $G$ be differentiable and strictly increasing entail that $c$ is the density of the transformed variables $(U,V):=(F(X),G(Y))$. Therefore, once one convinces oneself that $c_n(u,v)$ is simply the kernel density estimator of the bivariate density $c(u,v)$ of the pseudo-variables $(U,V)$, one directly draws its convergence properties by applying the results of the preceding subsection with $d=2$: 
\begin{lemma} \label{dens2}
For a choice of $a_n \simeq n^{-1/6}$,
for every $(u,v) \in (0,1)^2$, similar results of those of lemma \ref{dens1} hold for $\hat c_n$ with a rate of convergence of $n^{-1/3}$ and $(\ln\ln n/n)^{1/3}$ respectively.
  \end{lemma}

%------------------------------------------------------------
\subsection {An approximation lemma of $\hat c_n (u,v)$ by $c_n (u,v)$} \label{sectionapprox1}
The lemma of this section gives the rate of approximation of the kernel copula density estimator $\hat c_n (u,v)$ computed on the real data $(F_n(X_i), G_n(Y_i))$ by its analogue $c_n (u,v)$ computed on the pseudo-data $(U_i,V_i):=(F(X_i),G(Y_i))$. A similar result, but with a different proof, has been obtained in Fermanian \cite{F2005} theorem 1.

\begin{lemma} \label{approx1}
Let $(u,v) \in (0,1)^2$. If the kernel $K(u,v)=K_1(u)K_2(v)$ is twice differentiable with bounded second derivatives, then 
\begin{align*}
 |\hat c_n (u,v) - c_n (u,v)| &= o_P(a_n^2+1/\sqrt{na_n^2}) \\ 
 |\hat c_n (u,v) - c_n (u,v)| &= o_{a.s.} \left( {\sqrt {\frac{{\ln \ln n}}{{na_n^2 }}} } \right)  
\end{align*}
\end{lemma}

\textbf{Proof.}
We note $||.||$ a  norm for vectors. %To simplify notation on some occasions, we use Einstein's convention for $d$-variate vectors, i.e. writing $A$ instead of the sum of each coordinate $\sum_{j=1}^d {A_j}$. 
Set $
\Delta:=\hat c_n (u,v) - c_n (u,v) = \frac{1}{{na_n^2 }}\sum\limits_{i = 1}^n {\Delta_{i,n}(u,v) } 
$
with $$\Delta_{i,n}(u,v):=K\left( {\frac{{u - F_n (X_i )}}{{a_n }},\frac{{v - G_n (Y_i )}}{{a_n }}} \right) - K\left( {\frac{{u - F(X_i )}}{{a_n }},\frac{{v - G(Y_i )}}{{a_n }}} \right)$$
and define \[
Z_{i,n} : = \left( {\begin{array}{*{20}c}
   {F(X_i ) - F_n (X_i )}  \\
   {G(Y_i ) - G_n (Y_i )}  \\
\end{array}} \right).
\]
As mentioned in section \ref{approxglivenko}, $|F_n(X_i)-F(X_i)| \le ||F_n-F||_\infty$  and $|G_n(Y_i)-G(Y_i)| \le ||G_n-G||_\infty$ a.s. for every $i=1,\ldots,n$. Lemma \ref{glivenko} thus entails that the norm of $Z_{i,n}$ is independent of $i$ and such that 
\begin {align}
  ||Z_{i,n}|| &=O_P(1/\sqrt{n}) \mbox{ , } i=1,\ldots,n\label{KS} \\
  ||Z_{i,n}|| &=O_{a.s.}(\sqrt{\ln \ln n/n}) \mbox{ , } i=1,\ldots,n \label{chung}
\end {align}
   Now, for every fixed $(u,v) \in [0,1]^2$, since the kernel $K$ is twice differentiable, there exists, by Taylor expansion, random variables ${\tilde U_{i,n} }$ and ${\tilde V_{i,n} }$ such that, almost surely,
\begin{align*}
 \Delta &=  \frac{1}{{na_n^3 }}\sum\limits_{i = 1}^n {Z_{i,n}^T \nabla K\left( {\frac{{u - F(X_i )}}{{a_n }},\frac{{v - G(Y_i )}}{{a_n }}} \right)}  \\ 
  &+ \frac{1}{{2na_n^4 }}\sum\limits_{i = 1}^n {Z_{i,n}^T {\nabla ^2 K}\left( {\frac{{u - \tilde U_{i,n} }}{{a_n }},\frac{{v - \tilde V_{i,n} }}{{a_n }}} \right)Z_{i,n} } 
  :=\Delta_1+\Delta_2
  \end{align*}
%\begin{align*}
% \Delta_{i,n}(u,v) 
%  &= Z_{i,n}^T \frac{{\nabla K}}{{a_n }}\left( {\frac{{u - F(X_i )}}{{a_n }},\frac{{v - G(Y_i )}}{{a_n }}} \right)  \\
%  &+ Z_{i,n}^T \frac{{\nabla ^2 K}}{{2a_n^2 }}\left( {\frac{{u - \tilde U_{i,n} }}{{a_n }},\frac{{v - \tilde V_{i,n} }}{{a_n }}} \right)Z_{i,n}   
% \end{align*}
where $Z_{i,n}^T$ denotes the transpose of the vector $Z_{i,n}$ and $\nabla K$ and $\nabla^2 K$  the gradient and the Hessian respectively of the multivariate kernel function $K$
\begin{align*}
\nabla K = \left( {\begin{array}{*{20}c}
   {\frac{{\partial K}}{{\partial u}}}  \\
   {\frac{{\partial K}}{{\partial v}}}  \\
\end{array}} \right)
\mbox{ , }
 \nabla ^2 K = \left( {\begin{array}{*{20}c}
   {\frac{{\partial ^2 K}}{{\partial u^2 }}} & {\frac{{\partial ^2 K}}{{\partial u\partial v}}}  \\
   {\frac{{\partial ^2 K}}{{\partial u\partial v}}} & {\frac{{\partial ^2 K}}{{\partial v^2 }}}  \\
\end{array}} \right)
\end{align*}
\textit{Negligibility of $\Delta_2$}:
By the boundedness assumption on the second-order derivatives of the kernel, and equations \ref{KS} and \ref{chung},
\begin{align*}
 \Delta_2 = O_P\left({\frac{1}{na_n^4}}\right)\mbox{ and }
 \Delta_2 = O_{a.s.}\left({\frac{\ln \ln n}{na_n^4}}\right).
\end{align*}

\textit{Negligibility of  $\Delta_1$}: By centering at expectations,
\begin{align*}
 \Delta _1  &= \frac{1}{{na_n^3 }}\sum\limits_{i = 1}^n {Z_{i,n}^T \left( {\nabla K\left( {\frac{{u - F(X_i )}}{{a_n }},\ldots} \right) - E\nabla K\left( {\frac{{u - F(X_i )}}{{a_n }},\ldots} \right)} \right)}  \\ 
  &+ \frac{1}{{na_n^3 }}\sum\limits_{i = 1}^n {Z_{i,n}^T E\nabla K\left( {\frac{{u - F(X_i )}}{{a_n }},\frac{{v - G(Y_i )}}{{a_n }}} \right)}   
  := \Delta _{11}  + \Delta _{12}  \\ 
\end{align*}
\textit{Negligibility of  $\Delta_{12}$}: Bias results on the bivariate gradient kernel estimator  (See Scott \cite{S1992} chapter 6) entail that 
$$
E\nabla K\left( {\frac{{u - F(X_i )}}{{a_n }},\frac{{v - G(Y_i )}}{{a_n }}} \right) = a_n^3 \nabla c\left( {u,v} \right) + O(a_n^5 )
$$
Cauchy-Schwarz inequality yields that
 $$|\Delta_{12}|  \le \frac{n||Z_{i,n}||}{na_n^3}\left\|E\nabla K\left( {\frac{{u - F(X_i )}}{{a_n }},\frac{{v - G(Y_i )}}{{a_n }}} \right) \right\| $$
 In turn, with equations \ref{KS} and \ref{chung},
\begin{align*}
\Delta _{12}   =O_P(1/\sqrt{n}) \mbox{ and }
\Delta _{12}  =O_{a.s} (\sqrt{\ln \ln n/n}).
\end{align*}
\textit{Negligibility of  $\Delta_{11}$}: Set $A_i=\nabla K\left( {\frac{{u - F(X_i )}}{{a_n }}, \ldots } \right) - E\nabla K\left( {\frac{{u - F(X_i )}}{{a_n }}, \ldots } \right)$.
Then, 
$$|\Delta _{11} | \le \frac{||Z_{n}||}{{na_n^3 }}\sum\limits_{i = 1}^n { ||A_i||}$$  Boundedness assumption on the derivative of the kernel imply that $||A_i|| \le 2C$ a.s. We apply  Hoeffding inequality for independent, centered,   bounded by $M$, but non identically distributed random variables $(\eta_j)$ (e.g. see \cite{B1998}),
\begin{align}
P(\sum_{j=1}^n \eta_j > t) \le \exp \left({- \frac{t^2}{2nM^2} }\right).\label{bosqbernstein}
\end{align}
Here, for every $\epsilon>0$, with $M=2C$, $\eta_i=||A_i||-E||A_i||$, $t=\epsilon n^{1/2}(\ln \ln n)^{1/2}$, we get that
\[
P\left( {\sum\nolimits_{i = 1}^n {(||A_i||-E||A_i||) }  > \epsilon \sqrt {n\ln \ln n} } \right) \leqslant \exp \left(-\frac {\epsilon^2 \ln \ln n}{4M^2} \right) = \frac{1}
{{(\ln n)^{\delta}}}
\]
with a $\delta>0$ and where the r.h.s. goes to zero as $n \to \infty$. Therefore, $\sum\nolimits_{i = 1}^n {(||A_i||-E||A_i||)}=O_P(\sqrt {n\ln \ln n})$.

For the almost sure negligibility, we get similarly by inequality \ref{bosqbernstein} that, for every $\epsilon>0$, with $t=\epsilon n^{(1+\delta)/2}$ and $\delta >0$, 
 \[
P\left( {\sum\nolimits_{i = 1}^n {(||A_i|| -E||A_i||)}  > \epsilon n^{(1 + \delta )/2} } \right) \leqslant \exp \left( {\frac{{ - \epsilon^2 n^\delta }}
{{4M^2 }}} \right)
\]
and  the series on the r.h.s is convergent.
In turn, the Borell-Cantelli lemma imply that $\sum\nolimits_{i = 1}^n {(||A_i|| -E||A_i||)}=O_{a.s.}(n^{(1 + \delta )/2})$.

It remains to evaluate $E||A_i||$. First, we have that $E||A_i|| \le 2E||\nabla K((u-F(X_i))/a_n, \ldots)||$. Second, since $K$ is differentiable and of product form $K(u,v)$ $=K_1(u)K_2(v)$, each sub-kernel is of bounded variations and can be written as a difference of two monotone increasing functions. For example, set $K_1  = K_1^a  - K_1^b$ and define $
K^*  := (K_1^a  + K_1^b)K_2$.
 We have, \[
\left| {\frac{{\partial K}}
{{\partial u}}} \right| \leqslant \left( {|(K_1^a)'| + |(K_1^b)'|} \right)K_2  = ((K_1^a)' + (K_1^b)')K_2 : = \frac{{\partial K^* }}
{{\partial u}}
\]
where the equality proceeds from the positivity of the derivatives. As a consequence,  
$$E\left|\frac{\partial K}
{\partial u}((u-F(X_i))/a_n, \ldots)\right|\le E\frac{\partial K^*}
{\partial u}((u-F(X_i))/a_n, \ldots)$$
and similarly for the other partial derivative. The r.h.s. of the previous inequality is, after an integration by parts, of order $a_n^3$ by the results on the  kernel estimator of the gradient of the density (See Scott \cite{S1992} chapter 6). Therefore, $\sum\nolimits_{i = 1}^n {E||A_i||}=O(na_n^3)$.

Recollecting all elements, we eventually obtain that 
\begin{align*}
\Delta_{11} &= O_P \left( {\frac{{\sqrt {n\ln \ln n} +na_n^3}}
{{\sqrt n na_n^3 }}} \right) = O_P \left( {\frac{{\sqrt {\ln \ln n} }}
{{na_n^3 }}+\frac{1}{\sqrt n}} \right) = o_P \left( {\frac{1}
{{\sqrt {na_n^2 } }}} \right). \\
%Since $K$ is differentiable, it is of bounded variations and can be written as a difference of two monotone increasing functions. $K=K^B-K_B$. Therefore, since
%$E||A_i|| \le 2E||\nabla K((u-F(X_i))/a_n, \ldots)||$ and $||\nabla K||\le ||\nabla K^B||+||\nabla K_B||=\nabla K^B+\nabla K_B$, we get that
%$$E|A_i| \le 2E\nabla K^B((u-F(X_i))/a_n, \ldots)+2E_B\nabla K_B((u-F(X_i))/a_n, \ldots)$$
%and each term on the r.h.s. is of order $a_n^3$, by the results on the  kernel estimator of the gradient of the density (See Scott \cite{S1992} chapter 6). Therefore, $\sum\nolimits_{i = 1}^n {E|A_i|}=O(na_n^3)$, which, in turn entails that $$\Delta_{11} = O_P \left( {\frac{{\sqrt {n\ln \ln n} +na_n^3}}
%{{\sqrt n na_n^3 }}} \right) = O_P \left( {\frac{{\sqrt {\ln \ln n} }}
%{{na_n^3 }}+\frac{1}{\sqrt n}} \right) = o_P \left( {\frac{1}
%{{\sqrt {na_n^2 } }}} \right).
%$$
\Delta_{11} 
 &= O_{a.s.} \left( {\frac{{n^{(1 + \delta )/2} +na_n^3}}
{{na_n^3 }}\sqrt {\frac{{\ln \ln n}}
{n}} } \right) \\
&= O_{a.s.} \left( {\sqrt {\frac{{\ln \ln n}}{{na_n^2 }}} \frac{1}{{n^{(1 - \delta )/2} a_n^2 }} +\sqrt{\frac{\ln \ln n}{n}}} \right)
 = o_{a.s.} \left( {\sqrt {\frac{{\ln \ln n}}
{{na_n^2 }}} } \right)
\end{align*}
for $\delta$ small enough ($<1/3$ for $a_n \simeq n^{-1/6}$).
\qed

%By using uniform convergence properties of the kernel (derivative) estimators, one can extend the previous approximation lemma for fixed $(u,v)$ to the whole compact set $[0,1]^2$:
%\begin{corollary]
%\[
%\mathop {\sup }\limits_{(u,v) \in [0,1]^2 } |\hat c_n (u,v) - c_n (u,v)| = 
%o_P \left( {\sqrt {\frac{{\ln n}}{{na_n^2 }}} } \right)
%\]
%\[
%\mathop {\sup }\limits_{(u,v) \in [0,1]^2 } |\hat c_n (u,v) - c_n (u,v)| = 
%o_{a.s.}P \left( {\sqrt {\frac{{\ln \ln n}}{{na_n^2 }}} } \right)
%\]
%\end{corollary]
%\textbf{Proof.}
%Omitted for brevity. It follows the lines of the preceding lemma, with a chaining technique \textit{à la Kolmogorov }.
%\qed

%---------------------------------------------------------- 
\subsection{An approximation lemma for $\hat c_n (F_n (x),G_n (y))$ by $\hat c_n (F(x),G(y))$} \label{sectionapprox2}
The lemma of this subsection gives the rate of deviation of the kernel copula density estimator $\hat c_n $ from a varying location $(F_n(x), G_n(y))$ to a fixed location $(F(x),G(y))$.
\begin{lemma} \label{approx2}
With the same assumptions as in the preceding lemma, we have
\begin{align*} 
 \hat c_n (F_n (x),G_n (y)) - \hat c_n (F(x),G(y))&=  o_P \left( {a_n^2+\frac{1}{{\sqrt {na_n^2 } }}} \right) \\
 \hat c_n (F_n (x),G_n (y)) - \hat c_n (F(x),G(y)) &=   O_{a.s.} \left( {\sqrt{\frac{\ln \ln n}{n}}} \right)
 \end{align*}
\end{lemma}

\textbf{Proof.}
We proceed similarly as in the preceding lemma. Set
\begin{align}
 \Delta_n(x,y):= \hat c_n (F_n (x),G_n (y)) - \hat c_n (F(x),G(y))= \frac{1}{{na_n^2 }}\sum\limits_{i = 1}^n {\Delta '_{i,n} (x,y)} \label{def}
\end{align}
with  
\begin{align*}
\Delta '_{i,n} (x,y) &:=
K\left( {\frac{{F_n (x) - F_n (X_i )}}{{a_n }},\frac{{G_n (y) - G_n (Y_i )}}{{a_n }}} \right) \\
& - K\left( {\frac{{F(x) - F_n (X_i )}}{{a_n }},\frac{{G(y) - G_n (Y_i )}}{{a_n }}} \right)
\end{align*}
and define \[
Z_{n}(x,y) : = \left( {\begin{array}{*{20}c}
   {F_n(x ) - F (x )}  \\
   {G_n(y ) - G (y )}  \\
\end{array}} \right)
\]

We first express $\Delta '_{i,n} (x,y)$ at a fixed location $(F(x),G(y))$ by a Taylor expansion and by bounding uniformly the second order terms, 
\begin{align}
\Delta '_{i,n} (x,y) = Z_n^T (x,y)\frac{{\nabla K}}{{a_n }}\left( {\frac{{F(x) - F_n (X_i )}}{{a_n }},\frac{{G(y) - G_n (Y_i )}}{{a_n }}} \right) + \frac{||Z_n ||^2_\infty  }{a_n^2} R_1 \label{taylor1}
\end{align}
where $R_1$ is uniformly bounded almost surely: $R_1=O_{a.s.}(1)$.
We then go from the data $(F_n(X_i) , G_n(Y_i))$ to the pseudo but fixed w.r.t. $n$ data $(F(X_i) , G(Y_i))$. By a second Taylor expansion, 
\begin{align}
 &\frac{{\nabla K}}{{a_n }}\left( {\frac{{F(x) - F_n (X_i )}}{{a_n }},\frac{{G(y) - G_n (Y_i )}}{{a_n }}} \right) \nonumber \\ 
  &= \frac{{\nabla K}}{{a_n }}\left( {\frac{{F(x) - F(X_i )}}{{a_n }},\frac{{G(y) - G(Y_i )}}{{a_n }}} \right) \nonumber \\ 
  &+ Z_{i,n}^T \frac{{\nabla ^2 K}}{{2a_n^2 }}\left( {\frac{{F(x) - F(X_i )}}{{a_n }},\frac{{G(y) - G(Y_i )}}{{a_n }}} \right) + \frac{||Z_{n} ||_\infty}{a_n^2 }R_2 \label{taylor2}.
 \end{align} 
 where $R_2=o_{a.s.}(1)$ uniformly in $i$, $x$ and $y$.
Therefore, plugging \ref{taylor1} and \ref{taylor2} in \ref{def}, we get
\begin{align*} 
 \Delta _n (x,y)  &= \frac{{Z_n^T (x,y)}}{{na_n^2 }}\sum\limits_{i = 1}^n {\frac{{\nabla K}}{{a_n }}\left( {\frac{{F(x) - F(X_i )}}{{a_n }},\frac{{G(y) - G(Y_i )}}{{a_n }}} \right)}  \\ 
  &+ \frac{{Z_n^T (x,y)}}{{na_n^2 }}\sum\limits_{i = 1}^n {Z_{i,n}^t \frac{{\nabla ^2 K}}{{2a_n^2 }}\left( {\frac{{F(x) - F(X_i )}}{{a_n }},\frac{{G(y) - G(Y_i )}}{{a_n }}} \right)}  \\ 
  &+ R_3 \frac{{||Z_n ||_\infty ^2 }}{{a_n^4 }} .
 \end{align*}
with the remainder term $R_3=O_{a.s.}(1)$ uniformly.
As before, the  properties of the kernel (derivate) density estimator (See Scott \cite{S1992} chapter 6) entails that 
\begin{align*}
\frac{1}{{na_n^3 }}\sum\limits_{i = 1}^n {\nabla K\left( {\frac{{F(x) - F(X_i )}}{{a_n }},\frac{{G(y) - G(Y_i )}}{{a_n }}} \right)}  = O_P (a_n^2+1/\sqrt{na_n^4}) .
\end {align*}
Therefore, using \ref{KS} and bounding uniformly the Hessian, \ref{def} becomes
\begin{align*} 
 \Delta _n (x,y) &= O_P \left( {a_n^2||Z_n ||_\infty+\frac{{||Z_n ||_\infty  }}{{\sqrt {na_n^4 } }}} \right) + O_P \left( {\frac{{||Z_n ||_\infty ^2 }}{{a_n^4 }}} \right) \\
 &= o_P \left( {a_n^2+\frac{1}{{\sqrt {na_n^2 } }}} \right).
 \end{align*}

Similarly, one gets with \ref{chung} and the strong consistency of the estimator of the gradient of the density that
$
 \Delta _n (x,y)=   O_{a.s.} \left( {\sqrt{\frac{\ln \ln n}{n}}} \right).
$
\qed

\addcontentsline{toc}{chapter}{Bibliography} %pour mettre la biblio dans la table des matières
%\nocite{*}  %pour inclure toutes les références du fichier de biblio même celles non citées dans le document tex
\bibliographystyle{plain}

\bibliography{bibdensiteconditionnelle}

\def\cprime{$'$}
\begin{thebibliography}{10}

\bibitem{B1998}
D.~Bosq.
\newblock {\em Nonparametric statistics for stochastic processes}, volume 110
  of {\em Lecture Notes in Statistics}.
\newblock Springer-Verlag, New York, second edition, 1998.
\newblock Estimation and prediction.

\bibitem{C1999}
S.~X. Chen.
\newblock Beta kernel estimators for density functions.
\newblock {\em Comput. Statist. Data Anal.}, 31(2):131--145, 1999.

\bibitem{D1974}
P.~Deheuvels.
\newblock Conditions n\'ecessaires et suffisantes de convergence ponctuelle
  presque s\^ure et uniforme presque s\^ure des estimateurs de la densit\'e.
\newblock {\em C. R. Acad. Sci. Paris S\'er. A}, 278:1217--1220, 1974.

\bibitem{D1979}
P.~Deheuvels.
\newblock La fonction de d\'ependance empirique et ses propri\'et\'es. {U}n
  test non param\'etrique d'ind\'ependance.
\newblock {\em Acad. Roy. Belg. Bull. Cl. Sci. (5)}, 65(6):274--292, 1979.

\bibitem{D1981}
P.~Deheuvels.
\newblock A {K}olmogorov-{S}mirnov type test for independence and multivariate
  samples.
\newblock {\em Rev. Roumaine Math. Pures Appl.}, 26(2):213--226, 1981.

\bibitem{DL2001}
L.~Devroye and G.~Lugosi.
\newblock {\em Combinatorial methods in density estimation}.
\newblock Springer Series in Statistics. Springer-Verlag, New York, 2001.

\bibitem{FY2005}
J.~Fan and Q.~Yao.
\newblock {\em Nonlinear time series}.
\newblock Springer Series in Statistics. Springer-Verlag, New York, second
  edition, 2005.
\newblock Nonparametric and parametric methods.

\bibitem{FYT1996}
J.~Fan, Q.~Yao, and H.~Tong.
\newblock Estimation of conditional densities and sensitivity measures in
  nonlinear dynamical systems.
\newblock {\em Biometrika}, 83(1):189--206, 1996.

\bibitem{F2005}
J.-D. Fermanian.
\newblock Goodness-of-fit tests for copulas.
\newblock {\em J. Multivariate Anal.}, 95(1):119--152, 2005.

\bibitem{FS2003}
J.-D. Fermanian and Scaillet O.
\newblock Nonparametric estimation of copulas for time series.
\newblock {\em Journal of Risk}, 5(4):25--54, 2003.

\bibitem{FRW2004}
J.-D. Fermanian, D.~Radulovi{\'c}, and M.~Wegkamp.
\newblock Weak convergence of empirical copula processes.
\newblock {\em Bernoulli}, 10(5):847--860, 2004.

\bibitem{GM1979}
T.~Gasser and H.-G. M{\"u}ller.
\newblock Kernel estimation of regression functions.
\newblock In {\em Smoothing techniques for curve estimation (Proc. Workshop,
  Heidelberg, 1979)}, volume 757 of {\em Lecture Notes in Math.}, pages 23--68.
  Springer, Berlin, 1979.

\bibitem{GM1990}
I.~Gijbels and J.~Mielniczuk.
\newblock Estimating the density of a copula function.
\newblock {\em Comm. Statist. Theory Methods}, 19(2):445--464, 1990.

\bibitem{GHNS2007}
J.~Gustafsonn, M.~Hagmann, J.P. Nielsen, and O.~Scaillet.
\newblock Local transformation kernel density estimation of loss distributions.
\newblock {\em Forthcoming in Journal of Business and Economic Statistics},
  2007.

\bibitem{GK2007}
L.~Gy{\"o}rfi and M.~Kohler.
\newblock Nonparametric estimation of conditional distributions.
\newblock {\em IEEE Trans. Inform. Theory}, 53(5):1872--1879, 2007.

\bibitem{H2007}
P.~D. Hoff.
\newblock Extending the rank likelihood for semiparametric copula estimation.
\newblock {\em Annals Appl. Stats.}, 1(1):265--283, 2007.

\bibitem{HBG1996}
R.~J. Hyndman, D.~M. Bashtannyk, and G.~K. Grunwald.
\newblock Estimating and visualizing conditional densities.
\newblock {\em J. Comput. Graph. Statist.}, 5(4):315--336, 1996.

\bibitem{HY2002}
R.~J. Hyndman and Q.~Yao.
\newblock Nonparametric estimation and symmetry tests for conditional density
  functions.
\newblock {\em J. Nonparametr. Stat.}, 14(3):259--278, 2002.

\bibitem{J1997}
H.~Joe.
\newblock {\em Multivariate models and dependence concepts}, volume~73 of {\em
  Monographs on Statistics and Applied Probability}.
\newblock Chapman \& Hall, London, 1997.

\bibitem{L2007}
C.~Lacour.
\newblock Adaptive estimation of the transition density of a markov chain.
\newblock {\em Ann. Inst. H. Poincaré Probab. Statist.}, 43(5):571--597, 2007.

\bibitem{N1989}
{\`E}.~A. Nadaraya.
\newblock {\em Nonparametric estimation of probability densities and regression
  curves}, volume~20 of {\em Mathematics and its Applications (Soviet Series)}.
\newblock Kluwer Academic Publishers Group, Dordrecht, 1989.
\newblock Translated from the Russian by Samuel Kotz.

\bibitem{P1962}
E.~Parzen.
\newblock On estimation of a probability density function and mode.
\newblock {\em Ann. Math. Statist.}, 33:1065--1076, 1962.

\bibitem{PR1983}
B.~L.~S. Prakasa~Rao.
\newblock {\em Nonparametric functional estimation}.
\newblock Probability and Mathematical Statistics. Academic Press Inc.
  [Harcourt Brace Jovanovich Publishers], New York, 1983.

\bibitem{PC1972}
M.~B. Priestley and M.~T. Chao.
\newblock Non-parametric function fitting.
\newblock {\em J. Roy. Statist. Soc. Ser. B}, 34:385--392, 1972.

\bibitem{R1956}
M.~Rosenblatt.
\newblock Remarks on some nonparametric estimates of a density function.
\newblock {\em Ann. Math. Statist.}, 27:832--837, 1956.

\bibitem{R1969}
M.~Rosenblatt.
\newblock Conditional probability density and regression estimators.
\newblock In {\em Multivariate Analysis, II (Proc. Second Internat. Sympos.,
  Dayton, Ohio, 1968)}, pages 25--31. Academic Press, New York, 1969.

\bibitem{S1992}
D.~W. Scott.
\newblock {\em Multivariate density estimation}.
\newblock Wiley Series in Probability and Mathematical Statistics: Applied
  Probability and Statistics. John Wiley \& Sons Inc., New York, 1992.
\newblock Theory, practice, and visualization, A Wiley-Interscience
  Publication.

\bibitem{SW1986}
G.~R. Shorack and J.~A. Wellner.
\newblock {\em Empirical processes with applications to statistics}.
\newblock Wiley Series in Probability and Mathematical Statistics: Probability
  and Mathematical Statistics. John Wiley \& Sons Inc., New York, 1986.

\bibitem{S1959}
M.~Sklar.
\newblock Fonctions de r\'epartition \`a {$n$} dimensions et leurs marges.
\newblock {\em Publ. Inst. Statist. Univ. Paris}, 8:229--231, 1959.

\bibitem{S1980}
C.~J. Stone.
\newblock Optimal rates of convergence for nonparametric estimators.
\newblock {\em Ann. Statist.}, 8(6):1348--1360, 1980.

\bibitem{S1982}
W.~Stute.
\newblock A law of the logarithm for kernel density estimators.
\newblock {\em Ann. Probab.}, 10(2):414--422, 1982.

\bibitem{S1984}
W.~Stute.
\newblock Asymptotic normality of nearest neighbor regression function
  estimates.
\newblock {\em Ann. Statist.}, 12(3):917--926, 1984.

\bibitem{S1986a}
W.~Stute.
\newblock Conditional empirical processes.
\newblock {\em Ann. Statist.}, 14(2):638--647, 1986.

\bibitem{S1986b}
W.~Stute.
\newblock On almost sure convergence of conditional empirical distribution
  functions.
\newblock {\em Ann. Probab.}, 14(3):891--901, 1986.

\bibitem{VDV1998}
A.~W. van~der Vaart.
\newblock {\em Asymptotic statistics}, volume~3 of {\em Cambridge Series in
  Statistical and Probabilistic Mathematics}.
\newblock Cambridge University Press, Cambridge, 1998.

\bibitem{VDVW1996}
A.~W. van~der Vaart and J.~A. Wellner.
\newblock {\em Weak convergence and empirical processes}.
\newblock Springer Series in Statistics. Springer-Verlag, New York, 1996.
\newblock With applications to statistics.

\end{thebibliography}

%à rajouter dans la bibliographie

\end{document}